\documentstyle[preprint,prd,aps,floats]{revtex}
\begin{document}
\draft

\preprint{Imperial/TP/94-95/55}


\title{ALTERNATIVE DERIVATION OF THE HU-PAZ-ZHANG
MASTER EQUATION OF QUANTUM BROWNIAN MOTION }
\author{J. J. Halliwell\footnote{E-mail: j.halliwell@ic.ac.uk} and  T.
Yu\footnote{E-mail: ting.yu @ic.ac.uk}}
\address{Theoretical Physics Group, Blackett Laboratory, Imperial College,
London SW7 2BZ, U.K.}
\date{June, 1995}
\maketitle

\begin{abstract}
Hu, Paz and Zhang [ B.L. Hu, J.P. Paz and Y. Zhang, Phys. Rev. D
{\bf 45}  (1992) 2843] have derived an exact master equation for
quantum Brownian motion in a general environment via path integral
techniques. Their master equation provides a very useful tool to study
the decoherence of a quantum system due to the interaction with its
environment.
In this paper, we give an alternative and elementary derivation of the
Hu-Paz-Zhang master
equation, which involves tracing the evolution equation for the Wigner
function. We also discuss the master equation in some special
cases.

\end{abstract}



\section{INTRODUCTION}

Quantum Brownian motion (QBM) models provide a paradigm of
open quantum systems that has been very useful in quantum
measurement theory [1], quantum optics [2] and decoherence [3-5]. One of the
advantages of the QBM models is that they are reasonably simple, yet
sufficiently complex to manifest many important features of
realistic physical processes.

Central to the study of QBM is the master equation for the reduced
density operator of the Brownian particle, derived by tracing out
the environment in the evolution equation for the combined system
plus environment. A variety of such derivation have been given
[6-9]. The most general is that of Hu, Paz and Zhang [10,11], who used
path integral techniques and in particular, the Feynman-Vernon
influence functional.

The purpose of this paper is to provide an alternative and
elementary derivation of the Hu-Paz-Zhang master equation for QBM,
by tracing the evolution equation for the Wigner function of the
whole system.

\section{MASTER EQUATION FOR QUANTUM BROWNIAN MOTION}
The system we considered is a harmonic oscillator
with mass $M$ and bare frequency $\Omega$,  in interaction with a
thermal bath consisting of a set of harmonic
oscillators with mass $m_n$ and natural frequency $\omega_n$. The
Hamiltonian of the system plus environment is given by

\begin{equation} H={ p^2\over{2M}}+ {1\over2}  M\Omega^2 q^2+
\sum_n\left({p^2_n\over
{2m_n}}+ {1\over2}m_n\omega^2_n q^2_n\right)+q\sum_n C_nq_n \> ,\end{equation}
where $q, p$ and $ q_n,p_n$ are the coordinates and momenta of the Brownian
particle and oscillators, respectively, and $C_n$ are coupling constants.

The state of the combined system (1) is most completely described by a density
matrix $\rho (q,q_i;q',q'_i,t)$ where $q_i$ denotes $(q_1,... q_N)$, and
$\rho$ evolves according to
\begin{equation}
\dot\rho = -{i\over\hbar}[H\,, \rho] \>.\end{equation}

The state of the Brownian particle is described the reduced density
matrix,  defined by tracing over the environment,
\begin{equation}
\rho_r(q,q',t)=\int \prod_n (dq_ndq'_n\,\delta(q_n-q_n'))\rho
(q,q_i;q',q'_i,t) \>.
\end{equation}
The equation of time evolution for the reduced density matrix is
called the master equation. For a general environment,
Hu, Paz, and Zhang [10] derived the following master equation by using path
integral techniques:

\begin{eqnarray}
 i\hbar{ \partial\rho_r\over \partial t} &=&
- {\hbar^2\over 2M}\left({\partial^2\rho_r \over \partial
q^2} - {\partial^2\rho_r \over \partial q'^2 }\right) +
{1\over 2}M\Omega^2(q^2-q'^2)
\rho_r\nonumber  \\
& & + {1\over 2}M\delta\Omega^2(t)(q^2-q'^2)\rho_r\nonumber \\
& & - i\hbar\Gamma(t)
(q-q')\left( {\partial\rho_r\over \partial q} -{\partial\rho_r \over \partial
q'}\right) \nonumber  \\
& & - iM\Gamma(t)h(t)(q-q')^2\rho_r\nonumber \\
& & +\hbar \Gamma(t)f(t)(q-q')
\left({\partial\rho_r \over \partial q}+{\partial\rho_r\over\partial q'}\right)
\>.
\end{eqnarray}
The explicit form of the coefficients of the above equation will be
given later on.
The coefficient $\delta\Omega^2(t)$ is the frequency shift term, the
coefficients $\Gamma(t)$ is the ``quantum dissipative'' term, and the
coefficients
$\Gamma(t)h(t), \Gamma(t)f(t)$ are
``quantum diffusion'' terms. Generally, these coefficients are time dependent

and of quite complicated behaviour.

We find  it  convenient to use the Wigner function of the
reduced density matrix,
\begin{equation}\tilde W(q,p,t)={1\over {2\pi}}\int du\>e^{{iu p
/\hbar}}\rho_r\left(q-{u\over 2},q+{u\over 2},t\right)\>.
\end{equation}
Taking the Wigner transform
of (4), we obtain{\footnote {We believe Eq. (2.48) in Ref. [10]
contains some incorrect numerical factors.}}
\begin{eqnarray}
{\partial \tilde W\over \partial t}= &-&{1\over M} p{\partial\tilde W \over
\partial q } +M[\Omega^2 + \delta
   \Omega^2(t)]q{\partial\tilde W \over \partial p}\nonumber \\
&+& 2\Gamma(t){\partial (p\tilde W)\over \partial p}
    + \hbar M\Gamma(t)h(t){\partial^2\tilde W \over \partial
     p^2} \nonumber \\
&+& \hbar \Gamma(t)f(t) {\partial^2\tilde W\over {\partial q \partial
    p}}\>.
\end{eqnarray}
The inverse transformation of (5) is given by
\begin{equation}
\rho_r(q,q',t)=\int dp\> e^{-{ip(q-q')/\hbar}}\tilde W\left({{q+q'}\over
2},p,t\right).
\end{equation}

Our strategy for deriving the master equation (4) is to derive the
Fokker-Planck type equation (6) from the Wigner equation for the
total system. The master equation can be obtained from the Wigner
equation for the system by using  the transformation (7).

We shall make the following two assumptions:

(1) The system and the environment are initially uncorrelated, {\it {ie}}. the
initial Wigner function factors
\begin{equation}W_0 (q,p;q_i,p_i)=W^s_0(q,p) W^b_0(q_i,p_i)\>,\end{equation}
where $W^s_0$  and $W^b_0$  are the Wigner functions  of the system and the
bath, respectively, at $t=0$.

(2) The heat bath is initially in a thermal equilibrium state at
temperature $T=(k_B\beta)^{-1}$.
This means that the initial Wigner function of bath is of Gaussian form,
\begin{eqnarray}
W^b_0&=&\prod_n W^b_{n0}\nonumber \\
     &=&\prod_n N_n{\rm exp}\{-{2\over \omega_n\hbar}\tanh({1\over
2}\hbar \omega_n \beta)H_n\}\>,
\end{eqnarray}
where $H_n$ is the Hamiltonian of the $n$-th oscillator in the bath,
\begin{equation}
H_n={p^2_n\over 2m_n}+{1\over 2}m_n\omega_n^2q_n^2 \>.
\end{equation}
In addition, one easily  see that the initial moments of the bath are

\begin{eqnarray}
     & & \langle q_n(0)\rangle = \langle p_n(0) \rangle =0 \>,\\
     & &\langle q_n(0)q_m(0)\rangle =0\,\> ({\rm if}\> m \neq n)\>,\\
     & &\langle p_n(0)p_m(0)\rangle =0\,\> ({\rm if}\> m \neq n)\>,\\
     & &\langle q_n(0)p_m(0)+ p_m(0)q_n(0)\rangle =0 \>,
\end{eqnarray}
and
\begin{eqnarray}
\langle q^2_n(0)\rangle &
= & {\hbar\over 2m_n\omega_n}
\coth({1\over 2}\hbar\omega_n\beta)\> ,\nonumber \\
\langle p^2_n(0)\rangle & =
& {1\over 2}\hbar m_n\omega_n
\coth({1\over 2}\hbar\omega_n\beta) \>.
\end{eqnarray}

For the QBM problem described by (1) and (2), the Wigner function of the
combined  system plus  environment satisfies
\begin{eqnarray}
{\partial W \over \partial t}= & - & {p\over M}{\partial W\over
\partial q} + M\Omega^2 q{\partial W \over \partial p} \nonumber \\
                               & + & \sum_n \left (-{p_n \over
m_n}{\partial W \over \partial q_n}+m_n\omega_n^2 q_n{\partial W \over
\partial p_n}\right ) \nonumber \\
                               & + & \sum_nC_n\left (q_n{\partial
W\over \partial p}+q{\partial W \over \partial p_n}\right ) .\end{eqnarray}
By integrating over the bath variables on the both sides of the above
equation , one obtains
\begin{equation}{{\partial {\tilde W}}\over \partial t}=-{p\over M}{{\partial
{\tilde W}} \over \partial q}+M\Omega^2 q{{\partial {\tilde W}} \over
\partial p} +  \sum_n C_n\int \prod_i dq_idp_i  q_n{\partial
W\over\partial p} \>,\end{equation}
where $ {\tilde W}(q,p)$ is the reduced Wigner function and it follows
from Eq. (3) that

\begin{equation}{\tilde W }(q,p)=\int^{+\infty}_{-\infty} \prod_i dq_i
dp_i W(q,p;q_i,p_i)\> .
\end{equation}
This definition is equivalent to Eqs. (3) and (5). The first two
terms on the right-hand side of the Eq. (17) give rise to
the standard evolution equation of the system. The last term contains
all the
information about the behaviour of the system in the presence of
interaction with environment.

In what follows, we shall demonstrate that the quantity
\begin{equation}G(q,p)=\sum_n C_n \int \prod_i dq_i dp_i q_n W\end{equation}
appearing (differentiated with respect to $p$) in (17) can
be expressed in terms of $\tilde W $and its derivatives.
To this end, we first perform Fourier
transform  of $G(q,p)$
\begin{eqnarray}
G(k,k')&=&\int dqdp\,e^{ikq+ik'p}G(q,p)\nonumber \\
       &=&\sum_nC_n\int
          dqdp\prod_idq_idp_i\>q_n\,e^{ikq+ik'p}\,W(q,p;q_i,p_i)\>.
\end{eqnarray}

It is well known that $q(t),p(t)$ and $ q_n(t) , p_n(t)$ are related
to the classical evolution of their initial values $ q(0),p(0)$
and $q_n(0), p_n(0)$ through a
canonical transformation:

\begin{equation}{\bf z}(t)= U(t){\bf z}(0)\>,\end{equation}
where
$${\bf z}(t) =(q(t),q_1(t)...q_N(t);p(t),p_1(t),...p_N(t))\>. $$
Since the Hamiltonian (1) is quadratic, the Eq. (16) has the same
form as the classical Liouville equation, so the solution of
Eq. (16) is of the form,
\begin{equation}W_t({\bf z})=W_0(U^{-1}(t)\bf z) \>.\end{equation}
Changing the integration variables into their initial values by
this  canonical transformation, we obtain
\begin{eqnarray}
 G(k,k') &=& \int dq(0)dp(0)\prod_i dq_i(0)dp_i(0)\nonumber \\
         & & \times\left[fq(0)+gp(0)
               +\sum_n(f_nq_n(0)
              +g_np_n(0))\right]\nonumber \\
         & & \times {\rm exp}\left[ik\left(\alpha q(0)+\beta
              p(0))+\sum_n(a_nq_n(0)+b_np_n(0))\right)\right]\nonumber \\
         & & \times {\rm exp}\left[ik'M\left(\dot \alpha
             q(0)+\dot\beta p(0)+
             \sum_n(\dot a_n  q_n(0)+\dot b_n
              p_n(0))\right)\right]\nonumber \\
         & & \times W^s_0(q(0),p(0))W^b_0(q_i(0),p_i(0)\>.
\end{eqnarray}
Here the coefficients $f,g, f_n, g_n, \alpha, \beta, a_n, b_n$ are time
dependent. Their explicit values are not required.

Similarly, the Fourier transform of the reduced Wigner function is
\begin{eqnarray}
\tilde W(k,k')&=&\int dqdp\,e^{ikq+ik'p}\tilde W(q,p)\nonumber \\
              &=& \int dq(0)dp(0)\prod_i dq_i(0)dp_i(0)\nonumber \\
              & & \times {\rm exp}\left[ik\left(\alpha q(0)+\beta
                   p(0))+\sum_n(a_nq_n(0)+b_np_n(0))\right)\right]\nonumber \\
              & & \times {\rm exp}\left[ik'M\left(\dot \alpha
                   q(0)+\dot\beta p(0)+
                   \sum_n(\dot a_n  q_n(0)+\dot b_n
                    p_n(0))\right)\right]\nonumber \\
              & & \times W^s_0(q(0),p(0))W^b_0(q_i(0),p_i(0))\>.
\end{eqnarray}
Now compare $G(k,k')$ and $\tilde W(k,k')$. They differ by the terms
linear in $q(0), p(0), q_n(0), p_n(0)$ in the preexponential factor
in $G(k,k')$. Consider the factors $f_nq_n(0)$ and $g_np_n(0)$ in
$G(k,k')$. Since they multiply $W^b_0(q_i(0),p_i(0))$, and since
$W^b_0(q_i(0),p_i(0))$ is Gaussian in $q_n(0), p_n(0)$, the terms
$f_nq_n(0)W_0^b$ and $g_np_n(0)W^b_0$ may be replaced by terms of
the form ${\partial W^b_0 / \partial q_n(0)}, {\partial
W^b_0/\partial p_n(0)}$ up to time dependent factors. An integration
by parts then may be performed, and these factors are then
effectively  replaced by multiplicative factors of $k, k'$.

Similarly, the factors $fq(0), gp(0)$ in the prefactor in $G(k,k')$ may
be replaced by $\partial /\partial k, \partial /\partial k'$ (plus
some more factors of $k$ and $k'$). Hence, it is readily seen that
$G(k,k')$ is a linear combination of terms of the form $k, k',
\partial /\partial k, \partial /\partial k'$ operating on $\tilde
W(k,k')$, with time dependent coefficients.

Inverting the Fourier transform, it follows that

\begin{equation}G=A(t)q\tilde W + B(t)p\tilde W + C(t){\partial\tilde
W\over\partial q} + D(t){\partial\tilde
W \over \partial p}\> .\end{equation}
for some coefficients $A(t), B(t), C(t), D(t)$ to be determined. This
result  immediately leads to the general form Wigner equation :
\begin{eqnarray}
{\partial\tilde W \over\partial t}= & -
& {p\over M}{\partial\tilde W\over \partial
q} + M\Omega^2 q{\partial\tilde W \over \partial p}
+A(t)q{\partial\tilde W\over \partial p} \nonumber \\
                                    & + & B(t){\partial(p\tilde W) \over
\partial p} + C(t){\partial^2 \tilde W\over {\partial p\partial q}}
 + D(t){\partial^2 \tilde W\over{\partial p^2}}\>.
\end{eqnarray}

\section{DETERMINATION OF THE COEFFICIENTS(GENERAL CASE)}

Having found  the functional form of the Wigner equation (26) of
the Brownian particle, the next step is to determine the coefficients
in the equation. Undoubtedly, there is more than one way to do this.
Here we shall choose a way which is both mathematically simple
and physically heuristic. Towards this direction, let us
consider the time evolution of the expectation values of the
system variables: $ q, p, q^2, p^2 $ and ${1\over 2}(pq+qp)$.

By using Eq. (16), we have

\begin{eqnarray}
& &{d\over dt}\langle q\rangle  =  {1\over M}\langle p \rangle \>,\\
& &{d\over dt}\langle p\rangle  =  -M\Omega^2 \langle q\rangle -
   \sum_n C_n\langle q_n \rangle \>,\\
& &{d\over dt}\langle q^2 \rangle  =  {1\over M}\langle pq+qp \rangle \>, \\
& &{d\over dt}\langle p^2 \rangle  =  -M\Omega^2 \langle
    pq+qp \rangle
   -2\sum_n C_n \langle pq_n \rangle \>,\\
& &{d\over dt}\langle pq+qp \rangle  =  {2\over M}\langle p^2 \rangle - 2M
   \Omega^2 \langle q^2 \rangle - 2\sum_n C_n \langle qq_n \rangle \>.
\end{eqnarray}
Similarly, using Eq. (26) yields

\begin{eqnarray}
& &{d\over dt}\langle q\rangle  = {1\over M}\langle p \rangle \>,\\
& &{d\over dt}\langle p\rangle = -(M\Omega^2 + A)\langle q\rangle -
   B\langle p\rangle \>,\\
& &{d\over dt} \langle q^2 \rangle =  {1\over M}\langle pq+qp\rangle \>,\\
& &{d\over dt} \langle p^2 \rangle =  -(M\Omega^2 +
   A)\langle pq+qp \rangle - 2B\langle p^2 \rangle + 2D \>,\\
& &{d\over dt}\langle pq +qp\rangle  =
   {2\over M}\langle p^2 \rangle -2(M\Omega^2   +A)
   \langle q^2 \rangle - B\langle pq+qp \rangle + 2C \>.
\end{eqnarray}
Since the evolution equations of the expectation values are  confined to the
system variables, the above two sets of equations must be
identical.

Now by comparing (28) with (33) we see that
\begin{equation}
\sum_nC_n\langle q_n\rangle =A\langle q\rangle + B\langle p\rangle \>.
\end{equation}
Similarly, by comparing (31) with (36),  (30) with (35), respectively, we
get
\begin{eqnarray}
\sum_nC_n\langle qq_n \rangle &=& A\langle q^2\rangle +
                                  {B\over 2}\langle qp+pq\rangle -C \>,\\
\sum_nC_n\langle pq_n \rangle &=& {A\over 2}\langle pq+qp \rangle +
                                  B\langle p^2 \rangle - D \>.
\end{eqnarray}

The coefficients $A, B, C, D $ may now be determined from (37)-(39)
by regarding the expectation values $  \langle q \rangle , \langle
q_nq\rangle$ etc. as expectation values of Heisenberg picture
operators, and by solving the operator equation of motion. For
simplicity, we still use ordinary notation to represent an operator
without adding a hat on it.

The solution to the equation of motion may be written,
\begin{eqnarray}
q_n(t)&=&\alpha_nq(t)+\beta_n p(t)
         +\sum_m(a_{nm}q_m(0)+b_{nm}p_m(0)) \>,\\
q(t)  &=&\alpha q(0)+\beta p(0)
         +\sum_n(a_nq_n(0)+b_np_n(0)) \>,
\end{eqnarray}
for some time-dependent coefficients $\alpha_n, \beta_n, a_{nm},
b_{nm}, \alpha, \beta, a_n, b_n$.
Note that $q_n(t)$ has been expressed in terms of the final, not
initial values of $q, p$.
By substituting  Eq. (40) into Eq. (37), keeping (11) in mind, and
comparing the two sides of the resulting equation, we have
\begin{equation}
  A=\sum_nC_n\alpha_n\>,\>\> B=\sum_nC_n\beta_n \>.
\end{equation}
Similarly, substituting  (40) and (41)  into  (38) and (39), respectively,
we get
\begin{equation}
C=-\sum_{mn}C_n(a_{nm}a_m\langle q^2_m(0)\rangle +b_{nm}b_m\langle
p^2_m(0)\rangle )\>,
\end{equation}
\begin{equation}
D=-M\sum_{mn}C_n(a_{nm}\dot a_m\langle q^2_m(0)\rangle +b_{nm}\dot b_m\langle
p^2_m(0)\rangle)\>.
\end{equation}
Here we have made use of $p=M\dot q$. The coefficients $A, B, C, D$
are therefore completely determined by solving the equation of motion.
We now do this explicitly.

We have
\begin{equation} \ddot q(t)+ \Omega^2 q(t)=-{1\over M}\sum_n C_n
q_n(t) \>,\end{equation}
\begin{equation}\ddot q_n(t)+ \omega^2_n q_n(t)=-{C_n \over
m_n}q(t) \>.\end{equation}
The solution to Eq. (46) is as follows:
\begin{eqnarray}
q_n(t)&=& q_n(0)\cos(\omega_nt)+{p_n(0)\over m_n}{\sin(\omega_n t)\over
          \omega_n}\nonumber \\
      & & -C_n\int^t_0 ds\,{\sin[\omega_n(t-s)]\over
          \omega_n}{q(s)\over m_n}\>.
\end{eqnarray}
Combining (45) and (47) gives

\begin{equation}
\ddot q(t) + \Omega^2 q(t) +{2\over
M}\int_0^td\tau\eta(t-s)q(s)={f(t)\over M}\>,
\end{equation}
where
\begin{equation}
f(t)=-\sum_nC_n\left(q_n(0)\cos(\omega_nt) + {p_n(0)\over
m_n}{\sin(\omega_nt) \over \omega_n}\right).
\end{equation}
The kernel $\eta(s)$ is defined as
\begin{equation}
\eta(s) ={d\over ds}\gamma(s)\>,
\end{equation}
where
\begin{equation}
\gamma(s)=\int^{+\infty}_0d\omega\,{I(\omega)\over \omega}\cos(\omega s) \>.
\end{equation}
Here $I(\omega)$ is the spectral density of the environment:
\begin{equation}
I(\omega)=\sum_n\delta(\omega-\omega_n){C_n^2\over 2m_n\omega_n}\>.
\end{equation}

In order to get the expressions (40) and (41), we solve equation (48)
with
the following two different initial conditions:
\begin{equation}
q(s=0)=q(0)\>,\,\> \dot q(s=0)={p(0)\over M}\>.
\end{equation}
and
\begin{equation}
q(s=t)=q(t)\>,\,\> \dot q(s=t)={p(t)\over M}\>.
\end{equation}
where $t$ is any given time point. In doing so, we consider the
elementary functions $u_i(s) (i=1,2)$ introduced by
Hu, Paz and Zhang [10] which satisfy the following
homogeneous integro-differential equation

\begin{equation}{\ddot\Sigma}(s)+\Omega^2\Sigma(s)+{2\over M}\int_0
^s\,d\lambda\eta(s-\lambda)\Sigma(\lambda)=0
\end{equation}
with  the boundary conditions:
\begin{equation}u_1(s=0)=1\>,\,\,  u_1(s=t)=0\> ,
\end{equation}
and
\begin{equation}u_2(s=0)=0\>,\,\,  u_2(s=t)=1 \>.
\end{equation}

The solution to equation (55) with the initial condition (53) is
obtained as the linear combination of $u_1,u_2$,

\begin{equation}
w(s)=\left(u_1(s)-{\dot u_1(0)\over \dot
        u_2(0)}u_2(s)\right)q(0)+{u_2(s)\over \dot u_2(0)}\,{p(0)\over M}\>.
\end{equation}
The solution to equation (48) with the homogeneous initial conditions
can be formally written as
\begin{equation}
\tilde w(s)={1\over M}\int^s_0 d\tau G_1(s,\tau)f(\tau) \>.
\end{equation}
Where $G_1(s_1,s_2)$ is the Green function which can be constructed in
terms of $u_i(i=1,2)$:

\[G_1(s_1,s_2)= \left\{\begin{array}{ll}
                         {u_1(s_2)u_2(s_1)-u_2(s_2)u_1(s_1)\over
                          u_1(s_2)\dot u_2(s_2)-\dot u_1(s_2)u_2(s_2)}
                                 &\mbox{if $s_1>s_2$}\\
                          0      &\mbox{otherwise.}
                        \end{array}
                \right. \]
Note that $G_1(s_1,s_2)$ as the function of $s_1$
satisfies equation (55) with
\begin{equation}
G_1(s_1=s_2,s_2)= 0\>,\> \> {d\over ds_1}G_1(s_1=s_2,s_2)=1\>.
\end{equation}
Then the solution to the equation (48) with initial condition (53) reads
\begin{equation}
q(s)= w(s) + \tilde w(s),
\end{equation}
explicitly,
\begin{eqnarray}
q(s)&=&\left(u_1(s)-{\dot u_1(0)\over \dot
        u_2(0)}u_2(s)\right)q(0)+
        {u_2(s)\over \dot u_2(0)}\,p(0) \nonumber \\
    & &-\sum_n{C_n\over M}\int^s_0 d\tau\,G_1(s,\tau)\cos(\omega_n\tau)
         q_n(0)\nonumber \\
    & & - \sum_n{C_n\over M}\int^s_0d\tau\, G_1(s,\tau){\sin(\omega_n\tau)\over
        \omega_n}{p_n(0)\over m_n} .
\end{eqnarray}

It can be shown that the solution to the homogeneous equation (55)
with the initial conditions (54) is
\begin{equation}
u(s)=\left(u_2(s)-{\dot u_2(t)\over \dot u_1(t)}u_1(s)\right )q(t)+{u_1(s)\over
\dot u_1(t)}\,{p(t)\over M}\>,
\end{equation}
and
\begin{equation}
\tilde u(s) ={1\over M} \int^s_t d\tau G_2(s,\tau)f(\tau),\>(s\leq t)
\end{equation}
is the solution to the inhomogeneous equation (48) with the homogeneous
initial conditions
\begin{equation}
\tilde u(t)=0\>,\> \dot{\tilde u}(t)=0\>,
\end{equation}
where Green function $G_2(s_1,s_2)$ is
\[G_2(s_1,s_2)=\left\{\begin{array}{ll}
                      {u_1(s_2)u_2(s_1)-u_2(s_2)u_1(s_1)\over
                       u_1(s_2)\dot u_2(s_2)-\dot u_1(s_2)u_2(s_2)}
                              &\mbox{if $s_2>s_1$}\\
                       0      &\mbox{otherwise.}
                       \end{array}
               \right. \]
Hence, we get the solution to Eq. (48) with the initial conditions
(54)
\begin{eqnarray}
q(s)&=& u(s)+\tilde u(s)\nonumber \\
    &=& \left(u_2(s)-{\dot u_2(t)\over \dot u_1(t)}u_1(s)\right)q(t)+
        {u_1(s)\over  \dot u_1(t)}\,{p(t)\over M}\nonumber \\
    & & + \sum_n{C_n\over M}\int^t_sd\tau G_2(s,\tau)\cos(\omega_n\tau)\,
         q_n(0)\nonumber \\
    & & +\sum_n{C_n\over M}\int^t_sd\tau G_2(s,\tau){\sin(\omega_n\tau)\over
          \omega_n}{p_n(0)\over m_n} \>.
\end{eqnarray}
Substituting (66) into (47), one obtains
\begin{eqnarray}
q_n(t)&=&-{C_n\over
         m_n\omega_n}\int^t_0ds\,\sin[\omega_n(t-s)]\left(u_2(s)-
         {\dot u_2(t)\over \dot u_1(t)}u_1(s)\right )q(t)\nonumber \\
      & &-{C_n\over m_n\omega_n}\int^t_0ds\,\sin[\omega_n(t-s)]{u_1(s)\over
         \dot u_1(t)}\,{p(t)\over M}\nonumber \\
      & &+q_n(0)\cos(\omega_nt)+
         {p_n(0)\over m_n}{\sin(\omega_nt)\over\omega_n}\nonumber \\
      & &-{1\over M} \sum_m{C_nC_m\over m_n\omega_n}\int^t_0ds\int^t_sd\tau
         \sin[\omega_n(t-s)]G_2(s,\tau)\cos(\omega_m
          \tau)\,q_m(0)\nonumber \\
      & &-{1\over M} \sum_m{C_nC_m\over m_n\omega_n}\int^t_0ds\int^t_sd\tau
         \sin[\omega_n(t-s)]\,G_2(s,\tau){\sin(\omega_m\tau)\over
          m_m\omega_m}\>p_m(0) \>.
\end{eqnarray}
By using (42) we immediately arrive at
\begin{eqnarray}
A(t) &=&-\sum_n{C^2_n\over m_n\omega_n}\int^t_0ds\,\sin[\omega_n(t-s)]
         \left(u_2(s)-{\dot u_2(t)\over \dot u_1(t)}u_1(s)\right )\>, \\
B(t) &=&- {1\over M}\sum_n{C^2_n\over
         m_n\omega_n}\int^t_0ds\,\sin[\omega_n(t-s)]{u_1(s)\over\dot u_1(t)}\>
{}.
\end{eqnarray}
Furthermore, $A,B$ can be written as
\begin{eqnarray}
 A(t) &=& 2\int^t_0ds\eta(t-s)u_2(s)-2{\dot u_2(t)\over \dot u_1(t)}
          \int^t_0ds\eta(t-s)u_1(s)\>,\\
 B(t) &=& {2\over M\dot u_1(t)}\int^t_0ds\eta(t-s)u_1(s)\>.
\end{eqnarray}

{}From (62), the momentum of the Brownian particle is then
\begin{eqnarray}
p(s)&=&M\dot q(s)\nonumber \\
    &=&\left(\dot u_1(s)-{\dot u_1(0)\over \dot u_2(0)}\dot
       u_2(s)\right)\,Mq(0)+{\dot u_2(s)\over \dot
       u_2(0)}\,p(0)\nonumber \\
    & &-\sum_n\int^s_0d\tau
        G'_1(s,\tau)\cos(\omega_n\tau)\,q_n(0)\nonumber \\
    & &-\sum_n\int^s_0d\tau
        G'_1(s,\tau){\sin(\omega_n\tau)\over \omega_n}\,{p_n(0)\over m_n}\>.
\end{eqnarray}
Here ``prime'' stands for derivative with respect to the first variable
of $G_1(s,\tau)$.

With these results, It can be easily shown that
\begin{eqnarray}
C(t)&=&{\hbar\over M}\int^t_0\,d\lambda G_1(t,\lambda)\nu(t-\lambda)\nonumber
\\
    & & -{2\hbar\over M^2}\int^t_0ds\int^t_sd\tau \int^t_0d\lambda
       \eta(t-s)G_1(t,\lambda)G_2(s,\tau)\nu(\tau-\lambda)\>,
\end{eqnarray}
and
\begin{eqnarray}
D(t)&=&\hbar\int^t_0\,d\lambda G'_1(t,\lambda)\nu(t-\lambda)\nonumber \\
    & & -{2\hbar\over M}\int^t_0ds\int^t_sd\tau \int^t_0d\lambda
        \eta(t-s)G'_1(t,\lambda)G_2(s,\tau)\nu(\tau-\lambda)\>.
\end{eqnarray}
where $\nu(s)$ is defined as
\begin{equation}
\nu(s)=\int^{+\infty}_0d\omega I(\omega)\coth({1\over
2}\hbar\omega\beta)\cos(\omega
s)\>.
\end{equation}
It is seen that the coefficients $A(t), B(t), C(t), D(t)$ are dependent only on
the
kernels $\eta(s)$ and $\nu(s)$ and the initial state of the bath, not
dependent on the initial state of the system. Once the spectral density of the
environment is given, the elementary functions $u_i (i=1,2)$ can be
solved from equation (55), the Green functions $G_i (i=1,2)$ are then obtained.
Thus the coefficients of master equation can be
determined.

\section{PARTICULAR CASES}

In this section, we will consider some special cases. Let us at
first treat a special case in which we assume that the interaction
between the system and  environment is weak, so the $C_n$ are small.
In this case, the coefficients are of simple forms, and the
determination of these coefficients  is very simple and
straightforward. We shall work out these coefficients directly using
 the method in the last section, rather than the general formulae.

The solution to Eq. (46) may be written as
\begin{eqnarray}
q_n(t)& = & q_n(0) \cos (\omega_nt) +
{p_n(0)\over m_n}{\sin (\omega_nt) \over \omega_n}
\nonumber \\
&-& {C_n\over m_n} \int^t_0 dt'{\sin[\omega_n(t-t')] \over \omega_n}
    \cos[\Omega(t'-t)]\,q(t)\nonumber \\
&-& {C_n\over m_n} \int^t_0 dt'{\sin[\omega_n(t-t')] \over \omega_n}
     {\sin[\Omega(t'-t)]\over \Omega }\,{p(t)\over M} + O(C_n^2) \>.
\end{eqnarray}
Using Eq. (37) and  ignoring terms with higher than the second
order of $C_n$, we get
\begin{eqnarray}
\sum_nC_n \langle q_n(t) \rangle & = & \lbrace \sum_n
-{C_n^2\over m_n} \int^t_0 dt'
{\sin[\omega_n(t- t')]\over \omega_n}
\cos[\Omega (t'-t)] \rbrace \langle q(t) \rangle \nonumber \\
& + & \lbrace {1\over M}\sum_n -{C_n^2\over m_n}
\int^t_0 dt' {\sin [\omega (t-t')] \over
\omega_n}{\sin [\Omega (t'-t)]\over \Omega} \rbrace \langle p(t) \rangle \>.
\end{eqnarray}
Then we immediately get
\begin{equation} A(t)=2 \int^t_0 ds \eta (s) \cos (\Omega s) \>,
\end{equation}
\begin{equation}B(t)=-{2\over M\Omega}\int^t_0 ds \eta (s) \sin(\Omega s)\>
 .\end{equation}

We next evaluate  $\sum_n C_n \langle q(t)q_n(t) \rangle $ and $\sum_n
C_n\langle p(t)q_n(t)\rangle$. After a few
manipulations, we arrive at the expressions

\begin{equation}C(t)=- \sum_nC_n\lbrace \langle q(t)q_n(0)\rangle \cos
(\omega_nt) + \langle q(t)p_n(0)\rangle{\sin(\omega_nt)\over
m_n\omega_n}
\rbrace \>,
\end{equation}
and
\begin{equation}D(t)=- \sum_nC_n \lbrace \langle p(t)q_n(0)\rangle\cos
(\omega_nt)+ \langle p(t)p_n(0)\rangle {\sin(\omega_nt) \over
m_n\omega_n}\rbrace \>.
\end{equation}
To calculate $C(t)$ and $D(t)$, we need to expand $q(t)$ up to  the
second order of $C_n$:

\begin{eqnarray}
q(t)& = & q(0)\cos(\Omega t) + {p(0)\over M}{\sin(\Omega t)\over \Omega}
\nonumber \\
&-& \sum_n{C_n\over M}\int^t_0ds{\sin[\Omega (t-s)]\over
\Omega}\cos(\omega_ns)\>q_n(0)
\nonumber \\
&-& \sum_n{C_n\over M}\int^t_0ds{\sin[\Omega(t-s)]\over
\Omega}{\sin(\omega_ns)
\over \omega_n}\>{p_n(0)\over m_n} \nonumber \\
& + & O(C_n^2) \> .
\end{eqnarray}
The expansion of $p(t)$ is easily obtained from that of $q(t)$,
 \begin{equation}p(t)=M\dot q(t) .\end{equation}
With these results it is easy to compute $C(t)$ and $D(t)$:
\begin{equation}C(t)={\hbar\over
M\Omega}\int^t_0ds\nu(s)\sin(\Omega s) ,\end{equation}
\begin{equation}D(t)=\hbar\int^t_0ds\nu(s)\cos(\Omega s) .\end{equation}
This simple example exhibits the time dependency of the coefficients
of the master equation in a general environment. Eqs. (78), (79),
(84), (85) are in agreement with Hu, Paz and Zhang [10].

As another example, we briefly discuss the purly Ohmic case in the
Fokker-Planck limit (a particular high temperature limit), which has
been extensively discussed in the literature [6,10]. In this case  one has
\begin{eqnarray}
\eta(s-s')&=&M\gamma\delta'(s-s')\\
\nu(s-s') &=&{2M\gamma k_BT\over \hbar}\delta(s-s')
\end{eqnarray}
Then the equation (55) reduces to
\begin{equation}
\ddot u(s)+\Omega_{\rm ren}^2u(s)+\gamma\dot u(s)=-2\gamma\delta(s)u(0)
\end{equation}
where $\Omega^2_{\rm ren}=\Omega^2 -2\gamma\delta(0)$. After solving this
equation, a few calculations give
\begin{eqnarray}
A(t)&=&-2M\gamma\delta(0),\\
B(t)&=&2\gamma,\\
C(t)&=&0,\\
D(t)&=&2M\gamma k_BT,
\end{eqnarray}
then the Wigner  equation reads:
\begin{equation}
{\partial \over \partial t}\tilde W =-{p\over M}{\partial\tilde W
\over \partial
q}+M\Omega^2_{\rm ren}q{\partial\tilde W \over \partial
p}+2\gamma{\partial \tilde W\over \partial p} +2M\gamma
k_BT{\partial^2\tilde W\over
\partial p^2}\>.
\end{equation}
In this regime, the coefficients of this Wigner  equation are
constants.

\section*{Acknowledgements}
We are grateful to Roland Omn\`{e}s for suggesting this method of
deriving the master equation, and encouraging us to pursue it.


\begin{thebibliography}{99}

\bibitem{struct}W. Zurek, in {\it {Proceedings of the NATO Advanced
Study Institute on Non-Equilibrium Statistical Mechanics, Sante Fe,
1984}} (Plenum, New York, 1984).
\bibitem{struct}H. Carmichael, {\it {An Open Systems Approach to Quantum
Optics}} (Springer-Verlag, 1993).
\bibitem{struct}W.G. Unruh and W. Zureck, Phys. Rev. D {\bf 40}, 1071(1989).
\bibitem{struct}M. Gell-Mann and J. Hartle, Phys. Rev. D{\bf 47}, 3345(1993).
\bibitem{struct}H.F. Dowker and J.J. Halliwell,
Phys. Rev. D {\bf 46}, 1580(1992).
\bibitem{struct}A.O. Caldeira and A.J. Leggett,
Physica  {\bf 121A}, 587(1983).
\bibitem{struct}H. Dekker, Phys. Rev. {\bf A16}, 2116(1977).
\bibitem{struct}F. Haake and R. Reibold, Phys. Rev. A{\bf 32}, 2462(1985).

\bibitem{struct}J.P. Paz, in {\it {The Physical Origin of Time
Asymmetry}}, edited by J.J. Halliwell, J. Perez-Mercader,
and W. Zurek (Cambridge University
Press, Cambridge, 1994).



\bibitem{struct}B.L. Hu, J. Paz, and Y. Zhang, Phys.
Rev. D{\bf 45}, 2843(1992).
\bibitem{struct}B.L. Hu, J. Paz, and Y. Zhang, Phys.
Rev. D{\bf 47}, 1576(1993).



\end{thebibliography}
\end{document}